%% file: main.tex
\documentclass[]{spie}  

\usepackage{amsmath,amsfonts,amssymb}
\usepackage{graphicx}
\usepackage[colorlinks=true, allcolors=black]{hyperref}
\usepackage{pgfplots}
\pgfplotsset{compat=1.14}
\usepackage{enumitem}
\usepackage{siunitx}
\usepackage{subfig}
\usepackage{caption}
\usepackage{microtype}

\DeclareSIUnit{\um}{\micro\meter}

\title{Optics and broadband anti-reflection coatings\\ for the BA4-90/150 receiver}

\input{author_list_spie2026_ap}

\authorinfo{Further author information: (Send correspondence to A. R. Polish)\\E-mail: apolish@g.harvard.edu}

\begin{document} 
\maketitle

\begin{abstract}
The BICEP Array telescopes search for primordial B-mode polarization from inflationary gravitational waves. 
This signal is exceedingly faint, demanding excellent map depth and systematics control. 
The new BA4-90/150 receiver introduces a wide 80–\qty{169}{\giga\Hz} dichroic band, requiring upgrades throughout the optics chain to reduce loss, reflections, and thermal loading.
We developed improved anti-reflection (AR) coatings for our HMPE window, HDPE lenses, and nylon infrared filter, extending our AR technology to span more than an octave of bandwidth. 
The thermal filtering scheme and several mechanical elements were also updated to further suppress optical loss, reflections, and beam truncation. 
We aim to build on the proven success of deployed BICEP Array (BA) telescopes to produce a new small aperture instrument with the lowest optical systematics to date. 
\end{abstract}


\section{INTRODUCTION}
\label{sec:intro}  

The BICEP Array experiment is designed to make highly sensitive measurements of the polarization of the cosmic microwave background (CMB), with the primary goal of searching for the primordial B-mode signal generated by inflationary gravitational waves. 
Detecting this signal requires not only deep integration but also exceptional control of instrumental systematics. 
As map depth increases and statistical uncertainties continue to shrink, optical performance and systematics control become increasingly important factors in the overall sensitivity of the instrument.
Minimizing optical loss, reflections, scattering, thermal loading, and beam truncation is therefore essential for maximizing mapping speed while maintaining low systematic errors.

The new BA4-90/150 receiver has an 80–\qty{169}{\giga\Hz} dichroic band, the widest fractional bandwidth ever for a BICEP Array instrument. 
Achieving high optical efficiency across more than an octave of bandwidth presents significant challenges, particularly for anti-reflection (AR) coatings, where the limited availability of suitable low-loss, low-index dielectric materials poses an additional challenge. 
We are also motivated to update the thermal filtering strategy and several mechanical components to improve beam clearance and reduce thermal loading.

The BA4-90/150 receiver is scheduled to ship to the South Pole this season for deployment during the 2026--27 austral summer.
In these proceedings, we describe the principal optical design changes introduced for this receiver relative to previous BICEP Array instruments. 
We then present recent improvements to the fabrication of the thin HMPE vacuum windows.
Finally, we discuss the theory and practice of multi-layer AR coatings, including practical recipes for our polyethylene (PE) and nylon optics.

\section{RECEIVER OVERVIEW}

\begin{figure}[h]
    \centering
    \includegraphics[width=0.76\linewidth]{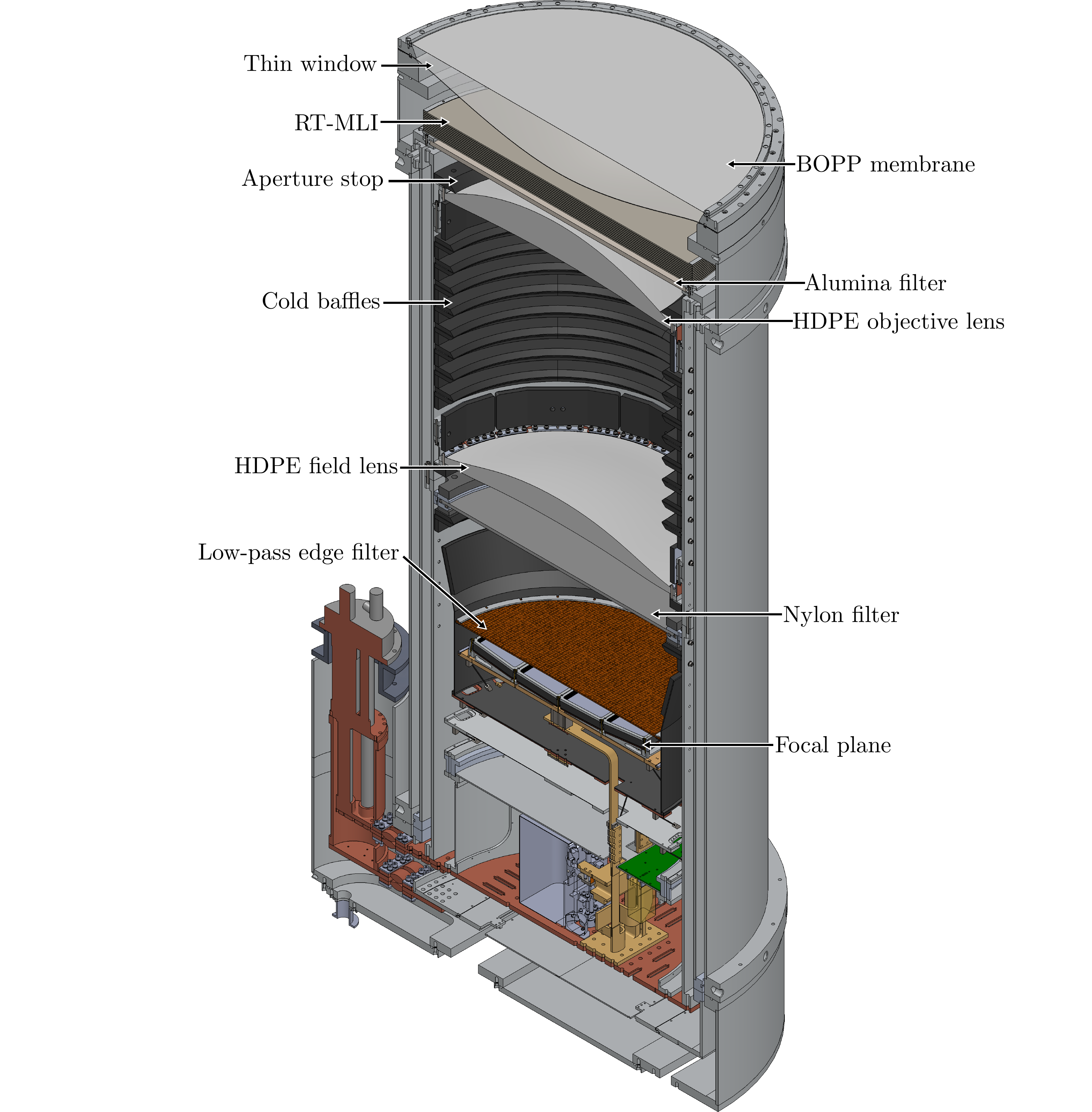}
    \caption{Preliminary cutaway drawing of BA4-90/150. The focal plane is inaccurate; this receiver will use hexagonal horn-coupled modules rather than the square antenna-coupled modules depicted here.}
    \label{fig:ba4_cutaway}
\end{figure}

The BICEP Array telescopes are small-aperture refractors optimized for low systematics and high sensitivity to the $\sim$\qty{1}{\degree} primordial B-mode signature.
Figure \ref{fig:ba4_cutaway} shows a cutaway diagram of BA4-90/150, illustrating the overall layout of the instrument.
The imaging optics consist of a pair of high-density polyethylene (HDPE) lenses held at \qty{4}{\kelvin}, whose prescription and design are detailed in an accompanying proceeding by Petroff et al.\footnote{Overview and status of BICEP Array's BA4-90/150 CMB polarimeter, Petroff et al., arXiv:2608.07817}
These lenses act to couple our focal plane of $\sim8000$ dichroic, dual-polarization horn-coupled detectors to the sky.
At the top of the cryostat is an ultra-thin vacuum window (see section \ref{sec:windows}), the only optic that must remain at \qty{300}{\K}.

Extensive optical filtering is required to mitigate potential above-band ``blue leak" sensitivity of the detectors and maintain the $\sim$\qty{300}{\milli\K} temperature of the focal plane.
From top to bottom, each BA receiver has a vacuum window, a radio-transparent multi-layer-insulation (RT-MLI) filter stack, a \qty{50}{\K} alumina filter, the \qty{4}{\K} HDPE objective and field lenses, a \qty{4}{\K} nylon filter, and finally a sub-K low-pass edge filter. 
Each optical element, except for the low-index RT-MLI foam, must be AR coated in order to minimize systematics. The AR coatings of the window, lenses, and nylon filter are detailed in section \ref{sec:ar}.

The optical design of the BA4-90/150 instrument largely follows that of previous BA receivers, as well as prior design work for the CMB-S4 small aperture telescopes \cite{presat_spie}. 
The remainder of this section discusses the changes to BA4-90/150 relative to other BA receivers, as well as the optical systematics motivating these changes.

\subsection{Motivation}
Although control of optical ghosting and large-angle response in the current BICEP Array receivers is generally good, these effects may still contribute to residual instrumental systematics. 
In particular, elevated noise observed in the BA2-150 receiver may be associated with bandpass mismatch, which can produce low-$\ell$ noise when mismatched detector pairs observe atmospheric emission.
Excess optical scattering can additionally lead to increased forebaffle coupling. 
Internal reflections, even when the fractional reflectivity of individual optical surfaces is low, can also produce complex systematic effects through multiple reflection paths. 

\begin{figure}[htbp]
\centering
\subfloat[BICEP3 full-instrument far sidelobe map \cite{christos_spie}\label{fig:1a}]{
    \includegraphics[width=0.42\textwidth]{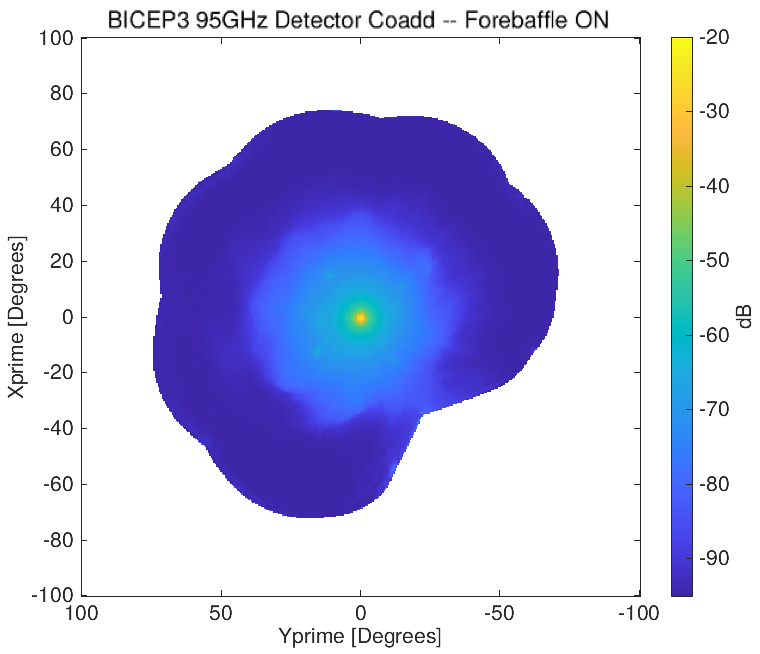}
}\hfill
\subfloat[BA3-220/270 single-detector near field beam map \cite{yuka_spie_2024}\label{fig:1b}] {
    \includegraphics[width=0.45\textwidth]{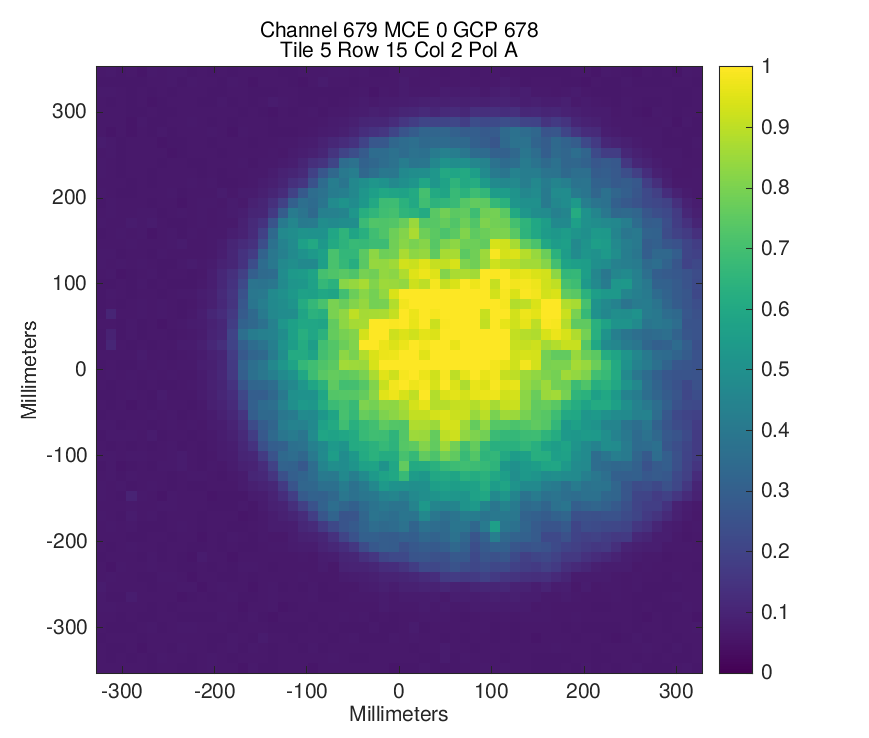}
}
\caption{Beam maps of BICEP3 (a) and BA3-220/270 (b), showing a variety of beam systematics including excess forebaffle coupling, four-fold features, and fringe effects.} \label{fig:scary_beams}
\end{figure}

Figure \ref{fig:scary_beams} shows a few of these features in the beams of BICEP3 and BA3-220/270. 
On the left, we can see a full-instrument far-sidelobe beam map of BICEP3, showing unintended four-fold features as well as significant coupling to the forebaffle (the broad ``shelf" out to around $35^\circ$) \cite{christos_spie}.
The forebaffle is deliberately absorbing, and we expect to see the edge of the forebaffle in our beams, but excess power within the forebaffle region but outside our main beam may indicate excess scattering somewhere in the optics chain.
The right subplot of Figure \ref{fig:scary_beams} is a single-detector near-field beam map on BA3-220/270, showing a ripple/fringe pattern that we do not currently understand, but which we suspect may be caused by internal reflections from some subset of our optics \cite{yuka_spie_2024}.

As the sensitivity of the BICEP program continues to improve, mitigating these optical systematics becomes increasingly important to ensure that instrument performance is not limited by residual optical effects.

\subsection{Optics upgrades}

BA4-90/150 is designed with a \qty{580}{\mm} aperture, increased from \qty{550}{\mm} on prior BICEP Array receivers.
This is only possible thanks to changes to optical elements to improve beam clearances:  
The window frame and clamp, which is responsible for securely clamping the slippery thin vacuum window under atmospheric pressure load, has been redesigned for better manufacturability and ease of use, as well as increased beam clearance. 
The alumina and nylon filters are also larger, again to improve beam clearance.
Finally, the RT-MLI filter stack now uses thinner metal spacer rings and mounts to \qty{50}{\K} rather than \qty{300}{\K}, which improves beam clearance and may slightly decrease thermal loading. 
In addition to the above changes driven by beam clearances, we will also implement improved baffles and blackening around the lens flexures and filter cells, ensuring that any stray light properly terminates on cold absorber.

\section{ULTRA-THIN VACUUM WINDOWS}
\label{sec:windows}

The telescope cryostat must have a vacuum window that can withstand atmospheric pressure over the full surface of the $\sim$\qty{75}{\cm} diameter window, yet still allow our detectors to make sensitive observations of the CMB. Low loss is especially critical in the vacuum window, as it is the only optical element at \qty{300}{\K}

\begin{figure}
    \centering
    \includegraphics[width=0.5\linewidth]{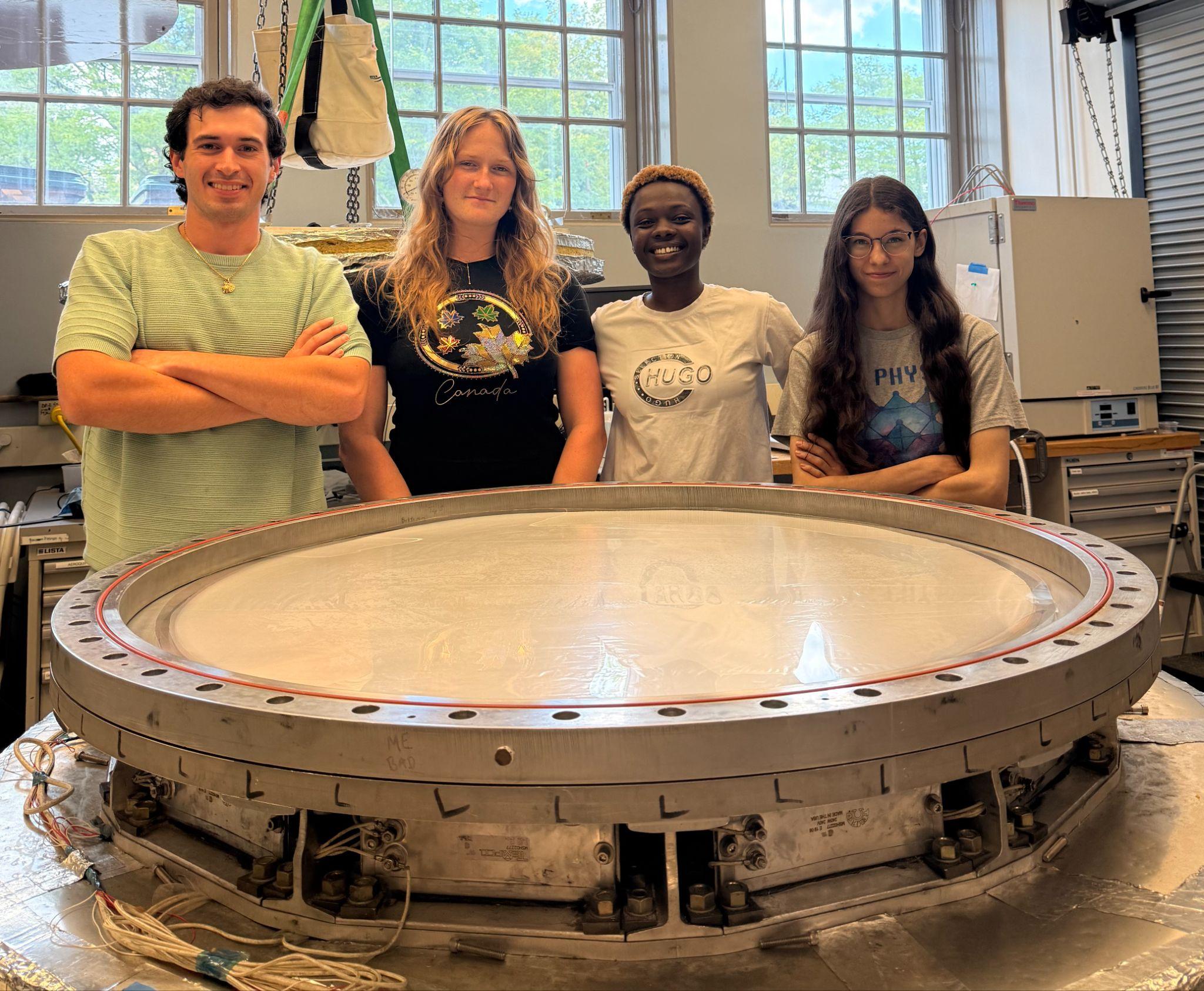}
    \caption{The bottom half of the large autoclave, with Summer 2026 window team for scale. From left to right, PJ Rioles, Sara Tomas, Faith Atieno, and Erin Cusson.}
    \label{fig:autoclave}
\end{figure}

Our ultra-thin vacuum windows are an in-house composite material, composed of woven high-modulus polyethylene (HMPE) fibers impregnated with low density polyethylene (LDPE). 
Careful temperature control is required to sufficiently melt the LDPE without damaging the HMPE fabric---we bake our windows at 135--\qty{137}{\celsius}. 
LDPE is highly viscous at these temperatures, so for good lamination we must also pressurize the layup to \qty{10}{atm} to force the LDPE into the weave \cite{window_paper}. 
This operation is performed in a custom-built autoclave, shown in Figure \ref{fig:autoclave}.

The completed windows are 1--2 mm thick (BA4-90/150's window will be \qty{1.33}{\mm} thick), and have been shown at full scale to have a mechanical safety factor better than 5.7. \cite{window_paper}.
We have successfully deployed these windows on BICEP3, BA2-150, and BA3-220/270, and will deploy another on BA4-90/150. 

\subsection{Process control}

When laminating a window, we typically place the window materials between two sheets of \SI{10}{mil} polyester as release layers, then put nonwoven PET breather cloth over the stack, and finally vacuum-bag the stack down to the bottom plate of the autoclave.\footnote{McMaster part 8567K94 (release layer), and FibreGlast products 579 (breather cloth) and 1688 (vacuum bag).}
The breather cloth ensures that we get good vacuum over the full area of the window, and should allow a good bake even if the vacuum bag develops a pinhole.
However, the breather cloth varies in thickness, and these thickness variations imprint into the thin window, even through the release layer.\footnote{This is unsurprising, as polyester is quite soft at \qty{135}{\celsius}.}
We believe this breather-cloth-induced lumpiness is the leading source of thickness variation across our thin windows, although variability of the input LDPE stock used for previous windows was also found to contribute significantly.

\begin{figure}
    \centering
    \includegraphics[width=0.7\linewidth]{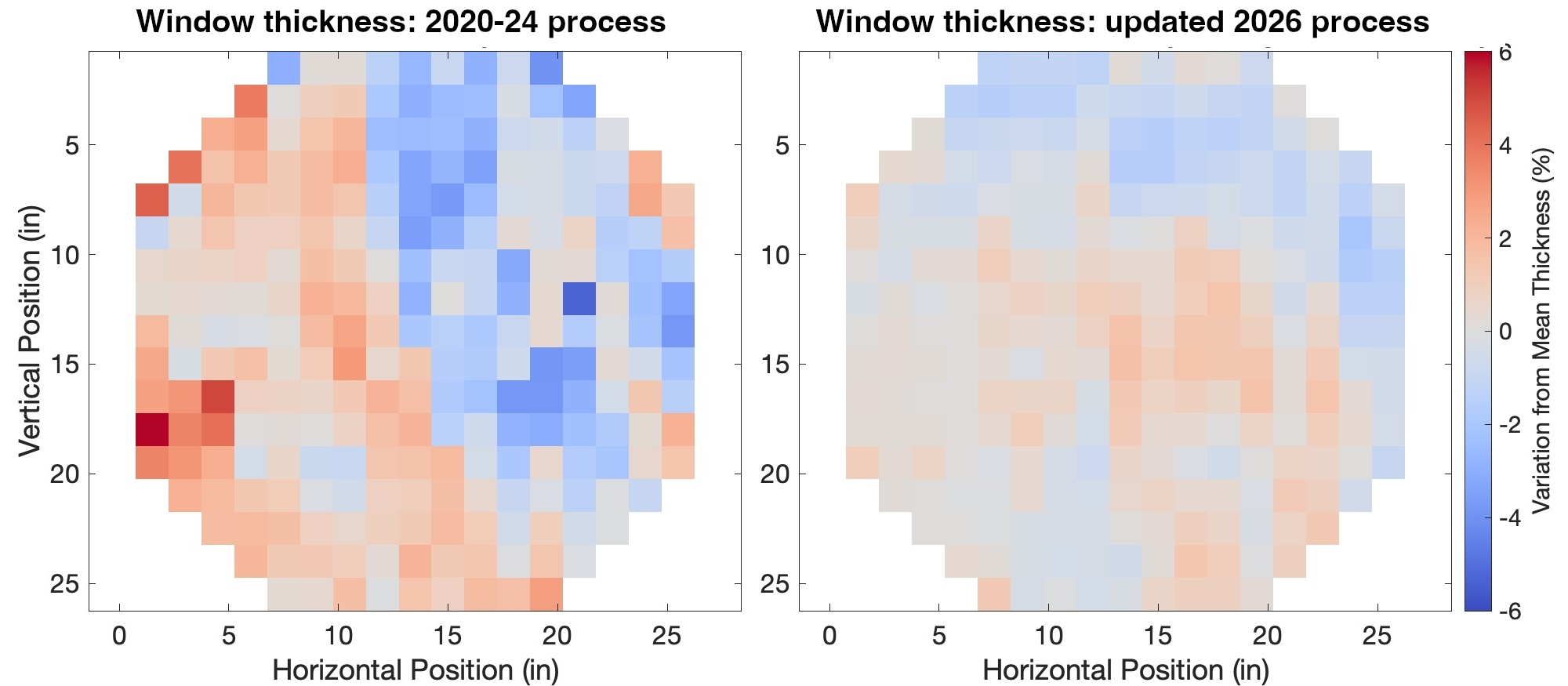}
    \caption{Thickness measurements of full-scale science-grade thin windows, showing improved thickness uniformity after introducing a caul plate.}
    \label{fig:thickness}
\end{figure}

On a representative thin window measuring \SI{1276}{\um} in thickness, the standard deviation of the thickness was \SI{24}{\um} (1.6\%), and the peak-to-peak variation was \SI{145}{\um} (11.4\%). 
The thickness variation of this window is shown on the left panel of Figure \ref{fig:thickness}.
We were motivated to improve our process for better repeatability and because, as we will discuss in section \ref{sec:ar}, the thickness of the window and other LDPE layers are crucial parameters in our AR prescriptions.

To improve our thin window process, we introduce a \SI{2}{\mm} thick sheet of 6061-T6 aluminum called a ``caul plate" into the stack, above the top release layer but below the breather cloth\cite{campbell_composites}.
This plate is intentionally fairly thin---a fully rigid plate would be physically unwieldy and introduce potential for significant thickness variation from non-parallelism between the caul plate and the oven's base plate.
The caul plate is, however, stiff enough to protect the window laminate from variations in the breather cloth, preventing imprinting.
It also somewhat mitigates variability in the input LDPE stock by locally smoothing out small-scale thickness differences.
We further mitigate LDPE stock variability by switching from fewer layers of thicker (4 and 6 mil) stock to many layers of a specific 2 mil stock that we have measured to have more uniform thickness.\footnote{Brand name Warp's Coverall, McMaster part 8651K92.}
Finally, we made improvements to the way that we cut and stack our HMPE and LDPE to improve cleanliness and allow us to thoroughly inspect the alignment of the layers before running the lamination bake.
The right panel of Figure \ref{fig:thickness} shows a representative thin window after our process improvements---it measures \SI{1213}{\um} thick, has a thickness standard deviation \SI{9}{\um} (0.7\%) and a peak-to-peak thickness variation of \SI{44}{\um} (3.6\%), a significant improvement over the previous methods.

Of the six full-scale bakes we have run on the improved process at the time of writing, five produced science-grade windows with similar thickness variability, with one window declared non-science-grade due to dust and oil inclusions. 
These results demonstrate that the revised fabrication process is robust, repeatable, and capable of consistently producing science-grade vacuum windows with excellent thickness control. 
The resulting manufacturing yield gives us confidence in the process for future production and, importantly, this process provides the thickness precision required to use LDPE as a predictable anti-reflection coating material in multi-layer wideband coating stacks.

\section{ANTI-REFLECTION COATINGS}
\label{sec:ar}

Anti-reflection coatings need to match the index of the optic to the index of free space, using destructive interference to minimize reflected power.

For an optic of index $n_c$ at wavelength $\lambda$, a single-layer AR coating should have index $n=\sqrt{n_c}$ and thickness $\lambda/4$.
However, broadband AR coatings are much more challenging, and require multiple layers. 
Each layer contributes a null, and one may choose how to arrange the nulls to optimize for different goals. 
In this case, we choose to minimize band-averaged reflection, so we use Chebyshev polynomials to determine the ideal index of each layer \cite{pozar}.

\begin{figure}
    \centering
    \includegraphics[width=0.6\linewidth]{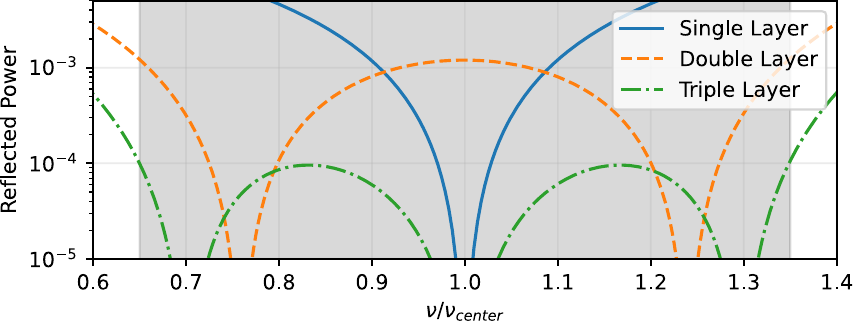}
    \caption{Reflections off ideal Chebyshev single, double, and triple layer AR coats for HDPE (n=1.55) and a fractional bandwidth of 0.7 \cite{miranda_spie_2024}.}
    \label{fig:chebyshev}
\end{figure}

BA4-90/150 will have a fractional bandwidth of 0.71, wider than any BICEP/Keck receiver to date. 
We require at least 2-layer AR coatings on all optics for $<1\%$ band-averaged reflection, and aim to fabricate AR recipes that provide $<0.1\%$ band-averaged reflection per surface.

\subsection{Practical AR materials}
Potential AR materials for plastic optics must be transparent to microwaves up to $\sim$\SI{300}{GHz},
have an index $<1.5$, must be available as sheets at least \SI{80}{cm} wide,
and must survive lamination at $\sim$135$^\circ$C.
This leaves a remarkably small list of candidate AR materials
\cite{miranda_spie_2024}. 

For BA4-90/150, we use two different formulations of PTFE: Expanded PTFE (ePTFE) from brands such as Teadit, DeWal, and Gore-Tex, is bulk PTFE that is biaxially stretched near its melting point. 
The resulting ePTFE material is highly compressible, allowing us to tune the combined index and thickness via controlled-pressure bakes in our autoclave. 
Sintered PTFE (sPTFE) from brands such as Fluorseals and Porex, begins as fine powder and is hot isostatic pressed at intermediate density. 
The resulting sPTFE material is largely incompressible, and generally higher index than ePTFE. 

Both of these materials are sold for industrial use, not precision optics, so we first characterize their index and loss using VNA reflectometry and quasi-optical cavity measurements \cite{brodi_spie_cavity}.

\subsection{Polyethylene AR coatings}

The lenses and thin window share a similar 2-layer AR prescription.
Due to the widely varying thickness of the lenses, we model their AR coats as single-sided coatings on semi-infinite slabs.
The window, on the other hand, is a single controllable thickness, which we adjust as a free parameter, effectively giving the thin-window coating an additional filter pole.
All AR recipes described in this and the following section were optimized using a global grid search over the space of all plausible material parameters, using very similar methodology to that used for prior AR recipes\cite{miranda_spie_2024}. 
The merit function used in the search is average in-band reflection, so if the material parameters are completely free, this optimization reduces to the ideal Chebyshev solution.
However, we are limited in the thicknesses and indices reachable with our materials, so a search is required to find the best option within the available parameter space.

The intermediate-index layer of both polyethylene AR coatings will be a custom sPTFE purchased from Fluorseals, in current production at the time of writing.
It will have a density of approximately 1.83 g/cc, leading to an index of approximately 1.35, and will be skived to a thickness of \qty{250}{\um}.
The lower-index AR layer will be an uncompressed sheet of nominally \qty{500}{\um} thick Teadit 24 SH (with a typical real thickness closer to \qty{560}{\um}).

\begin{figure}
    \centering
    \includegraphics[width=0.6\linewidth]{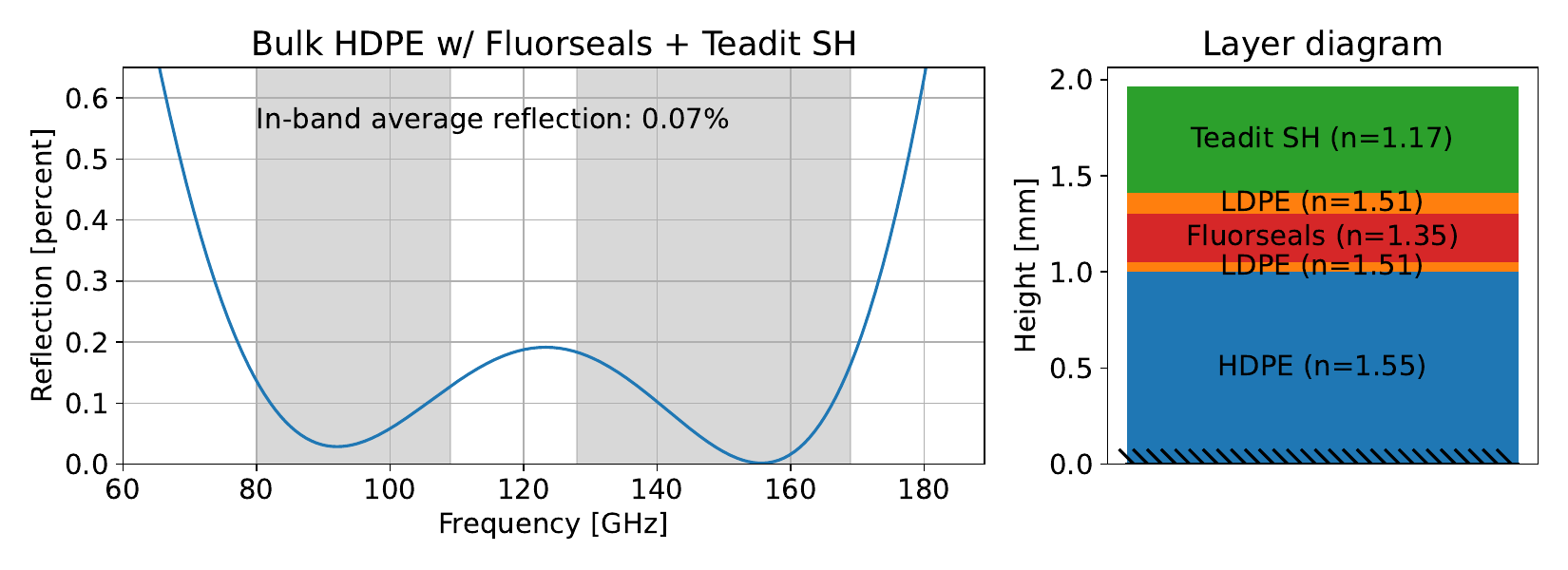}
    \caption{Left, projected single-surface AR performance for a 2-layer HDPE AR coating. The grey bars show the science bands of BA4-90/150. Right, schematic diagram of the AR layup.}
    \label{fig:hdpe_ar}
\end{figure}

The coating for the lenses will consist of \qty{51}{\um} of LDPE, \qty{250}{\um} of Fluorseals sPTFE, \qty{108}{\um} of LDPE, and finally a \qty{556}{\um} layer of uncompressed Teadit SH, for a single-surface band-averaged reflection of $0.07\%$. 
The free parameters tuned in this optimization were the thickness of the two LDPE layers and the pressure at which the Teadit SH is compressed.
Figure \ref{fig:hdpe_ar} shows the projected single-surface reflection spectrum and a visualization of the layer stackup. 

\begin{figure}[h]
    \centering
    \includegraphics[width=0.6\linewidth]{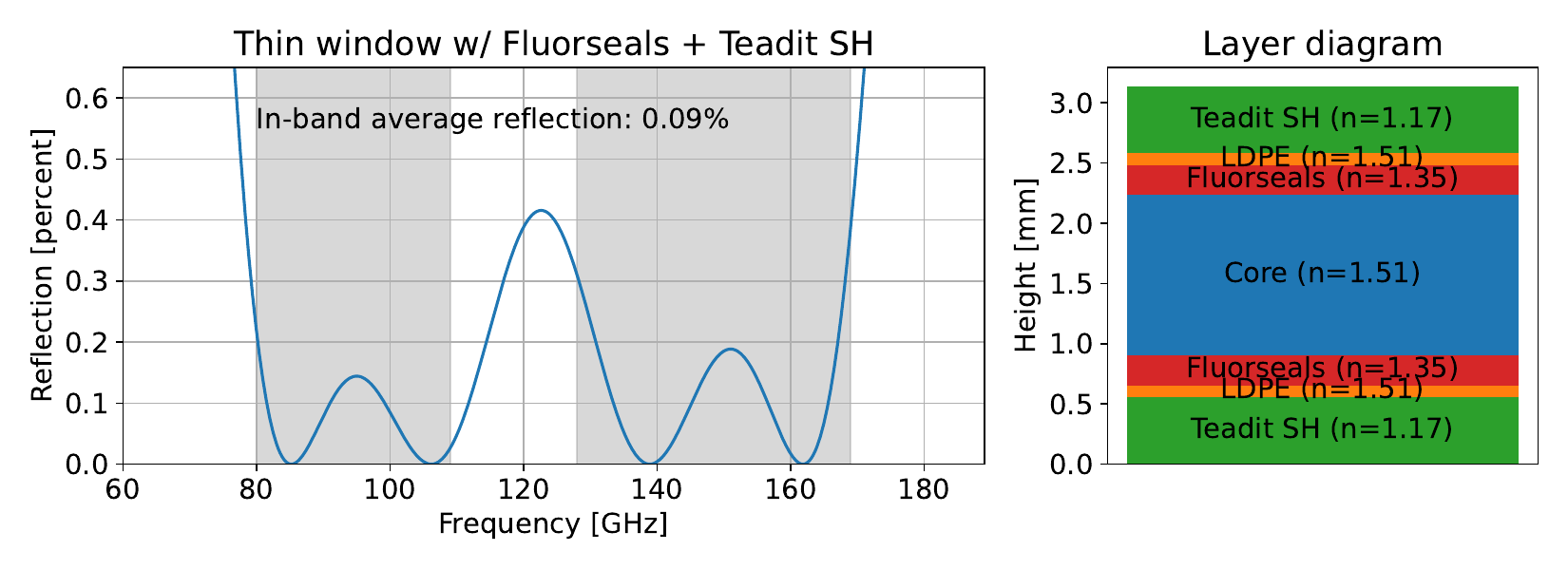}
    \caption{Left, projected AR performance for a 2-layer thin window AR coating. Right, schematic diagram of the AR layup.}
    \label{fig:window_ar}
\end{figure}

For the thin window, we allow the window thickness to be an additional free parameter that may range from \qty{1.2}{\mm} to \qty{2}{\mm}, accomplished in practice by putting more or fewer layers of LDPE into the core. 
The AR recipe will consist of a \qty{1.33}{\mm} core, \qty{250}{\um} of Fluorseals sPTFE, \qty{98}{\um} of LDPE, and finally a \qty{556}{\um} layer of uncompressed Teadit SH.
The full 2-sided AR-coated window should have a band-averaged reflection of $0.09\%$, and the reflection spectrum and layer stackup are shown in Figure \ref{fig:window_ar}.

\begin{figure}[h]
    \centering
    \includegraphics[width=0.6\linewidth]{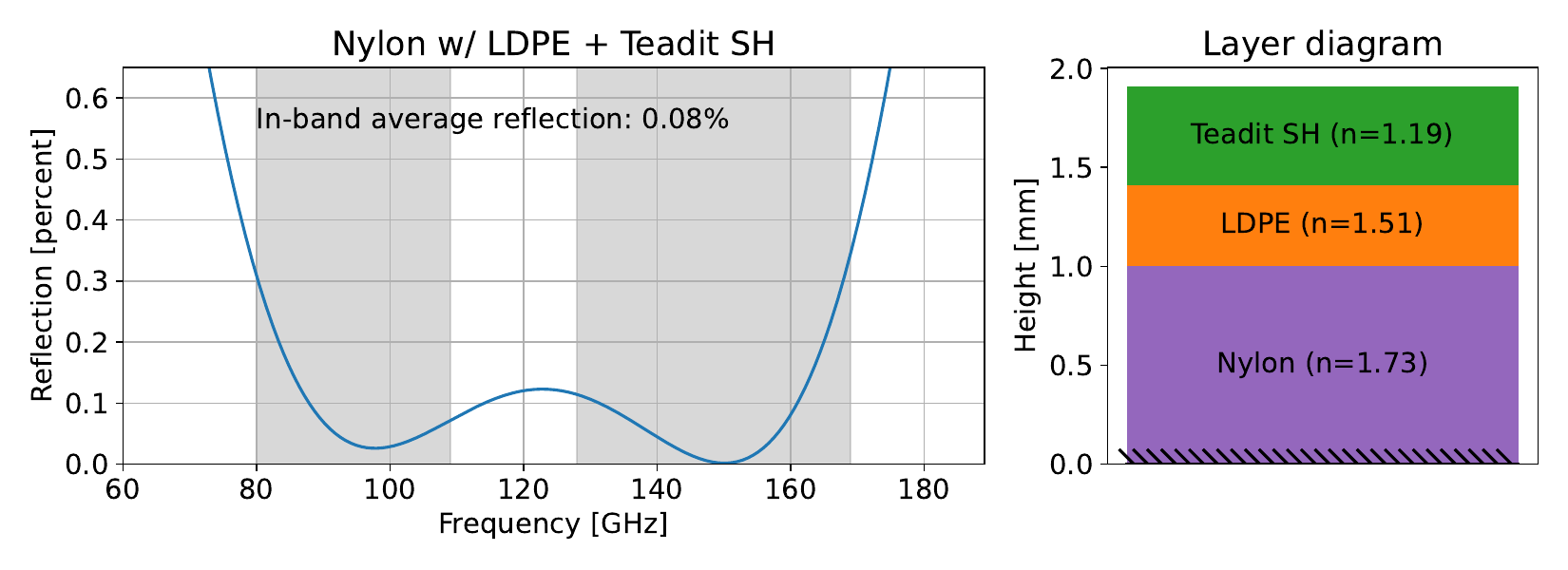}
    \caption{Left, projected single-surface AR performance for a 2-layer Nylon AR coating. Right, schematic diagram of the AR layup.}
    \label{fig:nylon_ar}
\end{figure}

\subsection{AR coatings for other optics}

The alumina filter, the highest-index optic in our telescope, 
will use a novel 3-layer AR coating currently under development.

Finally, the nylon filter will use LDPE as the mid-index AR layer, and slightly compressed Teadit SH as the lower-index AR layer.
Thick layers of LDPE are a surprisingly challenging AR material, since it is a viscous liquid at the temperatures of our bakes (unlike PTFEs, which have a much higher melting point). 
However, we are confident that the lessons learned from improving the window lamination process control will directly transfer to using LDPE as an AR layer.

The model AR recipe for the Nylon filter consists of \qty{408}{\um} of LDPE, followed by \qty{502}{\um} of Teadit SH that has been slightly compressed to an index of 1.19.
Figure \ref{fig:nylon_ar} shows the projected single-surface reflection and a visualization of the layer stackup. 

\begin{figure}[h]
    \centering
    \includegraphics[width=0.65\linewidth]{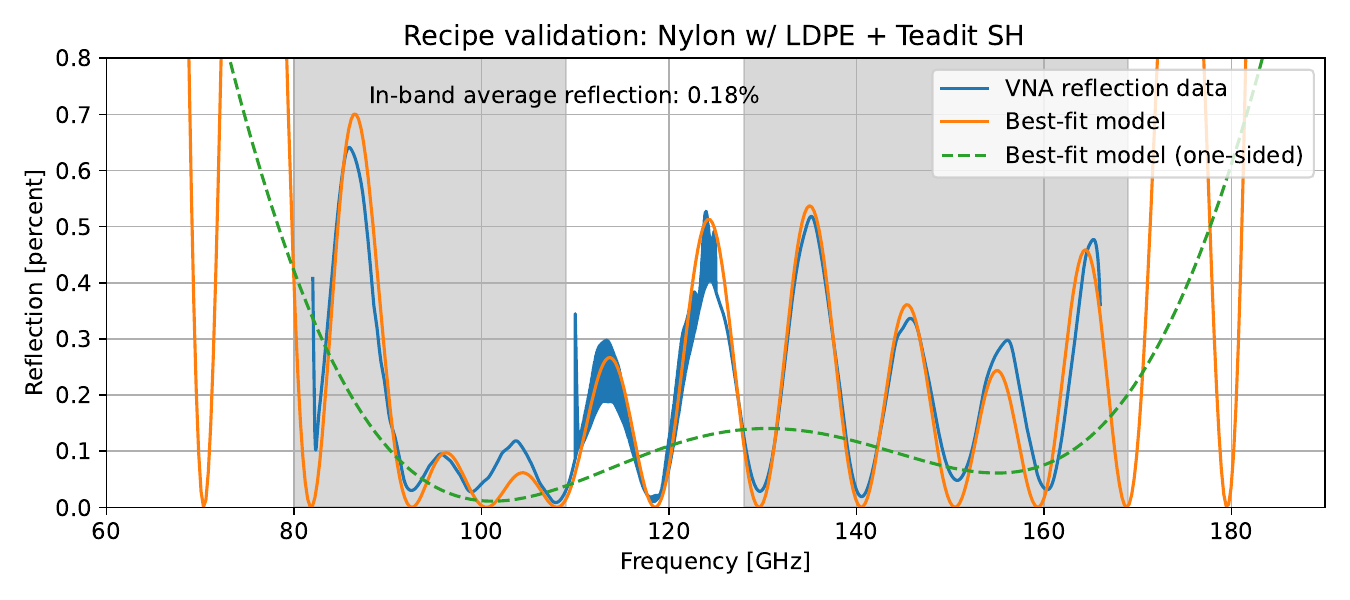}
    \caption{Measured performance of a prototype of the nylon filter AR recipe. The ``one-sided" spectrum was generated from the same fit parameter values as the two-sided best-fit model, but applied to one side of a semi-infinite slab. It has a band-averaged reflection of $0.11\%$, and is provided for comparison with  Figure \ref{fig:nylon_ar}.}
    \label{fig:nylon_ar_real}
\end{figure}

In order to validate the nylon AR recipe, we fabricated a small-scale test sample laminated onto a roughly quarter-inch-thick slab.
We then measure the reflected power off the sample using a VNA-based free-space reflectometry setup, using several frequency extender heads to cover nearly the full band of BA4-90/150.
Fitting to the measured data recovers an achieved AR recipe of \qty{376}{\um} of LDPE, followed by \qty{495}{\um} of Teadit SH compressed to an index of 1.198. 
This tells us that we used slightly less LDPE and compressed the Teadit slightly more than the optimized recipe did, providing immediate guidance on how we might change our process variables to produce a more optimal AR coating. 
However, the as-built sample is very close to the optimal recipe (see Figure \ref{fig:nylon_ar}), with a band-averaged reflection of $0.18\%$, exceeding our target of $<0.1\%$ per-surface reflection. 
This test sample gives us confidence that the nylon AR recipe is correct and that we are able to produce it with real materials, so at the time of writing we are working to fabricate this coating at full scale and apply it to a science-grade nylon filter.

\section{CONCLUSIONS AND OUTLOOK}

The optics for BA4-90/150 follow proven heritage from BA receivers, with many incremental improvements, plus new multi-layer AR coatings to cover our widest band to date.
We have demonstrated a robust and repeatable fabrication process for thin HMPE vacuum windows with high science-grade yield, and developed practical multi-layer anti-reflection coating recipes for the polyethylene and nylon optical elements that provide broadband performance across the 90/150 GHz dichroic band. 
Measurements of prototype coated optics are in good agreement with the model expectations, providing confidence in the fabrication approach.
Optical elements are currently in production and will deploy with BA4-90/150 in the 2026--27 austral summer season.

\acknowledgments 
 
We acknowledge the National Science Foundation Division of Astronomical Sciences for their support of PreSAT under Grant Number 2216223. We also thank the Summer 2026 Window Team---Faith Atieno, Erin Cusson, August Sununu, and Sara Tomas---for their tireless efforts on fabricating our thin windows and AR coatings.

\bibliography{report} 
\bibliographystyle{spiebib} 

\end{document}

%% file: author_list_spie2026_ap.tex
\author[a,b]{\href{https://orcid.org/0000-0002-7822-6179}{A.~R.~Polish}}%
\author[c]{P.~A.~R.~Ade}%
\author[d,e]{\href{https://orcid.org/0000-0002-9957-448X}{Z.~Ahmed}}%
\author[f]{\href{https://orcid.org/0000-0001-6523-9029}{M.~Amiri}}%
\author[a]{\href{https://orcid.org/0000-0002-8971-1954}{D.~Barkats}}%
\author[g]{\href{https://orcid.org/0000-0002-3351-3078}{R.~Basu~Thakur}}%
\author[h]{\href{https://orcid.org/0000-0001-9185-6514}{C.~A.~Bischoff}}%
\author[i]{\href{https://orcid.org/0000-0003-0848-2756}{D.~Beck}}%
\author[g,j]{J.~J.~Bock}%
\author[k]{V.~Buza}%
\author[i,d]{\href{https://orcid.org/0000-0003-4541-7080}{B.~Cantrall}}%
\author[g]{\href{https://orcid.org/0000-0002-1630-7854}{J.~R.~Cheshire~IV}}%
\author[l]{J.~Connors}%
\author[m]{\href{https://orcid.org/0000-0002-2088-7345}{J.~Cornelison}}%
\author[n]{M.~Crumrine}%
\author[g]{A.~J.~Cukierman}%
\author[l]{E.~Denison}%
\author[o]{L.~Duband}%
\author[a]{\href{https://orcid.org/0000-0002-7059-8728}{M.~A.~Echter}}%
\author[p]{\href{https://orcid.org/0009-0007-6718-1730}{M.~Eiben}}%
\author[a,b]{\href{https://orcid.org/0000-0003-4117-6822}{B.~D.~Elwood}}%
\author[g]{\href{https://orcid.org/0000-0002-3790-7314}{S.~Fatigoni}}%
\author[q]{\href{https://orcid.org/0000-0001-8217-6832}{J.~P.~Filippini}}%
\author[i]{A.~Fortes}%
\author[g]{M.~Gao}%
\author[h]{C.~Giannakopoulos}%
\author[i]{N.~Goeckner-Wald}%
\author[i]{\href{https://orcid.org/0000-0001-5268-8423}{D.~C.~Goldfinger}}%
\author[r,s]{S.~Gratton}%
\author[i]{J.~A.~Grayson}%
\author[g]{\href{https://orcid.org/0009-0003-6999-0129}{A.~Greathouse}}%
\author[a]{\href{https://orcid.org/0000-0001-9292-6297}{P.~K.~Grimes}}%
\author[f]{M.~Halpern}%
\author[d,e]{S.~Henderson}%
\author[n]{\href{https://orcid.org/0000-0002-3437-5228}{T.~D.~Hoang}}%
\author[l]{J.~Hubmayr}%
\author[g]{\href{https://orcid.org/0000-0001-5812-1903}{H.~Hui}}%
\author[i]{K.~D.~Irwin}%
\author[t]{M.~Izquierdo~Poza}%
\author[g]{\href{https://orcid.org/0000-0002-3470-2954}{J.~H.~Kang}}%
\author[t]{\href{https://orcid.org/0000-0002-5215-6993}{K.~S.~Karkare}}%
\author[g]{S.~Kefeli}%
\author[a,b]{\href{https://orcid.org/0009-0003-5432-7180}{J.~M.~Kovac}}%
\author[i]{C.~Kuo}%
\author[n,u]{\href{https://orcid.org/0000-0002-4540-1495}{K.~Lasko}}%
\author[g]{\href{https://orcid.org/0000-0002-6445-2407}{K.~Lau}}%
\author[h]{M.~Lautzenhiser}%
\author[i]{\href{https://orcid.org/0000-0001-5677-5188}{T.~Liu}}%
\author[k,v]{\href{https://orcid.org/0000-0002-1414-7236}{S.~C.~Mackey}}%
\author[n]{N.~Maher}%
\author[j]{K.~G.~Megerian}%
\author[g]{L.~Minutolo}%
\author[g]{\href{https://orcid.org/0000-0002-4242-3015}{L.~Moncelsi}}%
\author[i]{Y.~Nakato}%
\author[g,j]{H.~T.~Nguyen}%
\author[g,j]{R.~O’Brient}%
\author[a]{S.~N.~Paine}%
\author[g]{A.~Patel}%
\author[a]{\href{https://orcid.org/0000-0002-4436-4215}{M.~A.~Petroff}}%
\author[o]{T.~Prouve}%
\author[n]{\href{https://orcid.org/0000-0003-3983-6668}{C.~Pryke}}%
\author[l]{C.~D.~Reintsema}%
\author[a]{P.~J.~Rioles}%
\author[g]{T.~Romand}%
\author[i]{M.~Salatino}%
\author[g]{A.~Schillaci}%
\author[a]{B.~Schmitt}%
\author[n,u]{\href{https://orcid.org/0000-0001-7387-0881}{B.~Singari}}%
\author[g,j]{A.~Soliman}%
\author[a]{T.~St.~Germaine}%
\author[g]{\href{https://orcid.org/0000-0003-0260-605X}{A.~Steiger}}%
\author[g]{B.~Steinbach}%
\author[c]{R.~Sudiwala}%
\author[i,d]{K.~L.~Thompson}%
\author[c]{\href{https://orcid.org/0000-0002-1851-3918}{C.~Tucker}}%
\author[j]{A.~D.~Turner}%
\author[w]{\href{https://orcid.org/0000-0002-3942-1609}{C.~Verg\`{e}s}}%
\author[k,v]{A.~G.~Vieregg}%
\author[g]{\href{https://orcid.org/0000-0002-8232-7343}{A.~Wandui}}%
\author[j]{A.~C.~Weber}%
\author[n]{\href{https://orcid.org/0000-0002-6452-4693}{J.~Willmert}}%
\author[g,d,e]{\href{https://orcid.org/0000-0001-5411-6920}{W.~L.~K.~Wu}}%
\author[i]{H.~Yang}%
\author[k,m]{\href{https://orcid.org/0000-0002-8542-232X}{C.~Yu}}%
\author[a]{\href{https://orcid.org/0000-0001-6924-9072}{L.~Zeng}}%
\author[d]{\href{https://orcid.org/0000-0001-8288-5823}{C.~Zhang}}%
\author[g]{S.~Zhang}%
\affil[a]{Center for Astrophysics, Harvard \& Smithsonian, Cambridge, MA 02138, USA}%
\affil[b]{Department of Physics, Harvard University, Cambridge, MA 02138, USA}%
\affil[c]{School of Physics and Astronomy, Cardiff University, Cardiff, CF24 3AA, UK}%
\affil[d]{Kavli Institute for Particle Astrophysics and Cosmology, Stanford University, Stanford, CA 94305, USA}%
\affil[e]{SLAC National Accelerator Laboratory, Menlo Park, CA 94025, USA}%
\affil[f]{Department of Physics and Astronomy, University of British Columbia, Vancouver, BC, V6T 1Z1, Canada}%
\affil[g]{Department of Physics, California Institute of Technology, Pasadena, CA 91125, USA}%
\affil[h]{Department of Physics, University of Cincinnati, Cincinnati, OH 45221, USA}%
\affil[i]{Department of Physics, Stanford University, Stanford, CA 94305, USA}%
\affil[j]{Jet Propulsion Laboratory, California Institute of Technology, Pasadena, CA 91109, USA}%
\affil[k]{Kavli Institute for Cosmological Physics, University of Chicago, Chicago, IL 60637, USA}%
\affil[l]{National Institute of Standards and Technology, Boulder, CO 80305, USA}%
\affil[m]{High-Energy Physics Division, Argonne National Laboratory, Lemont, IL, 60439, USA}%
\affil[n]{School of Physics and Astronomy, University of Minnesota, Minneapolis, MN 55455, USA}%
\affil[o]{Service des Basses Temp\'eratures, Commissariat \`a l'\'Energie Atomique, 38054 Grenoble, France}%
\affil[p]{Faculty of Physical Sciences, University of Iceland, 102 Reykjav\'ik, Iceland}%
\affil[q]{Department of Physics, University of Illinois at Urbana-Champaign, Urbana, IL 61801, USA}%
\affil[r]{Centre for Theoretical Cosmology, DAMTP, University of Cambridge, Cambridge CB3 0WA, UK}%
\affil[s]{Kavli Institute for Cosmology Cambridge, Cambridge CB3 0HA, UK}%
\affil[t]{Department of Physics, Boston University, Boston, MA 02215, USA}%
\affil[u]{Minnesota Institute for Astrophysics, University of Minnesota, Minneapolis, MN 55455, USA}%
\affil[v]{Department of Physics, University of Chicago, Chicago, IL 60637, USA}%
\affil[w]{Lawrence Berkeley National Laboratory, Berkeley, CA 94720, USA}